\documentclass[amssymb,prb,twocolumn,showpacs]{revtex4}
\usepackage{epsfig}
\usepackage{dcolumn}
\usepackage{amsmath}
\begin{document}
\title{\bf Effect of a in-plane magnetic field on the microwave assisted magnetotransport
in a two-dimensional electron system}
\author{Jes\'us I\~narrea$^{1,2}$ and Gloria Platero$^2$}
 \affiliation {$^1$Escuela Polit\'ecnica
Superior,Universidad Carlos III,Leganes,Madrid,Spain and  \\
$^2$Unidad Asociada al Instituto de Ciencia de Materiales, CSIC,
Cantoblanco,Madrid,28049,Spain.}
\date{\today}
%%%%%%%%%%%%%%%%%%%%%%%%%%%%%%%%%%%%%%%%%%%%%%%%%%%%%%%%%%%%%%%%%%
%%%%%%%%%%%%
\begin{abstract}
In this work we present a theoretical approach  to study the effect
of an in-plane (parallel) magnetic field on the microwave-assisted
transport properties of a two-dimensional electron system. Previous
experimental evidences show that microwave-induced resistance
oscillations and zero resistance states are differently affected
depending on the experimental set-up: two magnetic fields (two-axis
magnet) or one tilted magnetic field. In the first case, experiments
report a clear quenching of resistance oscillations and zero
resistance states. In a tilted field, one obtains oscillations
displacement and quenching but the latter is unbalanced and less
intense. In our theoretical proposal we explain these results in
terms of the microwave-driven harmonic motion performed by the
electronic orbits and how this motion is increasingly damped by the
in-plane field.

\end{abstract}
%%%%%%%%%%%%%%%%%%%%%%%%%%%%%%%%%%%%%%%%%%%%%%%%%%%%%%%%%%%%%%%%%%
%%%%%%%%%%%%
\maketitle

%\section{Introduction}
In recent years, with the rise of
Nanotechnology, a lot of effort has been devoted to the study and
research of physics of nano-devices, from theoretical, experimental
and application perspectives. In particular the response of such
systems to external, time-dependent or stationary, fields is
receiving much attention from the scientific community\cite{ina}.
%The potential applications of these results to design and build a
%broad spectrum of nanoscale devices will become increasingly
%important in the forthcoming years.
A remarkable example is the recently obtained microwave- (MW-)
induced resistance oscillations (MIRO) and zero resistance states
(ZRS)
\begin{figure}
\centering\epsfxsize=2.5in \epsfysize=3.0in
\epsffile{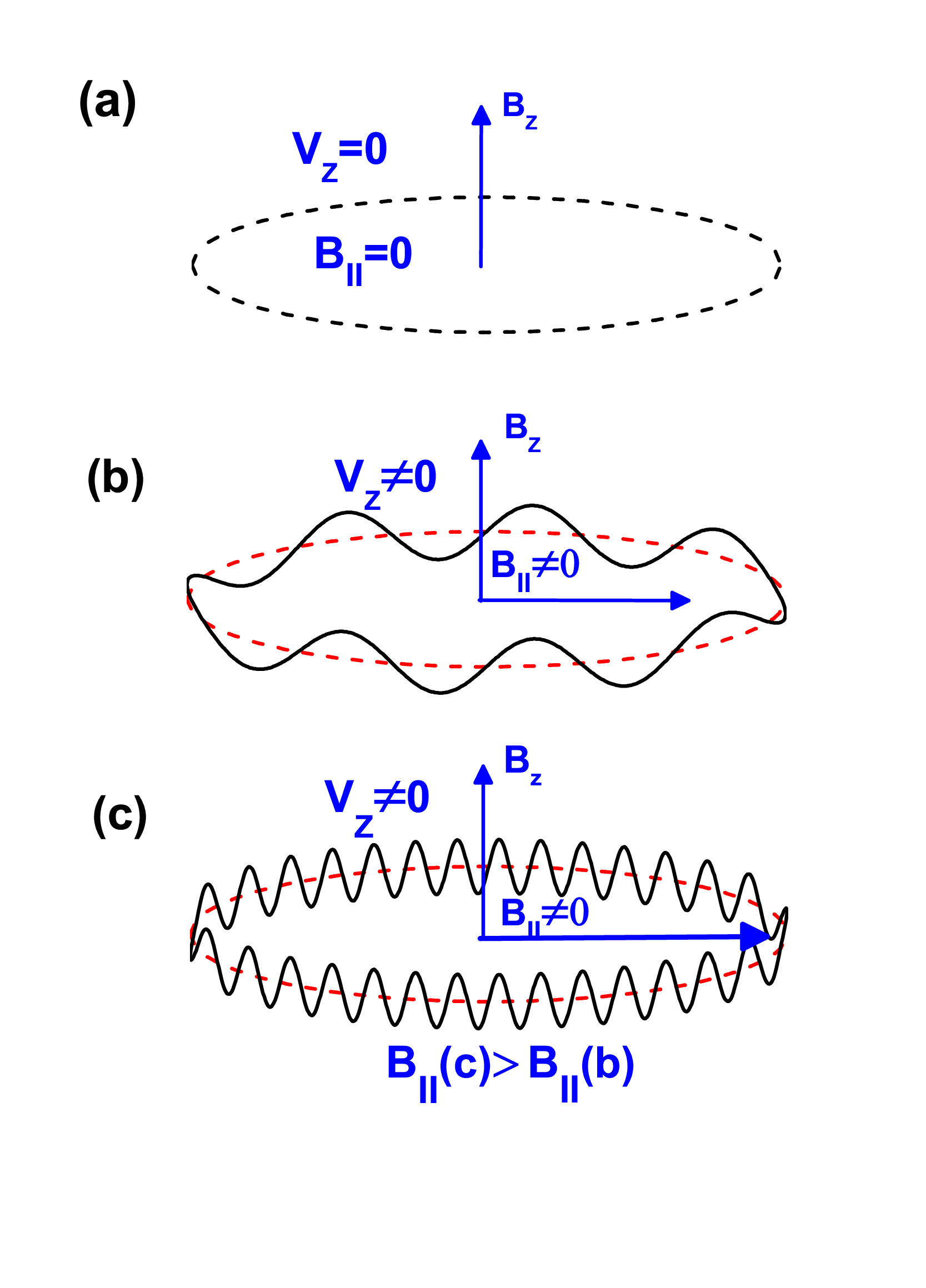} \caption{Schematic diagrams showing
the semiclassical description of electron trajectories in 2D systems
under different potentials and fields. 1(a): 2D ($x-y$ plane)
parabolic potential. 
%1 (b): 2D + 1D ($z$-direction ) parabolic
%potentials. 
1(b) and 1(c): 2D + 1D parabolic potentials and an
in-plane magnetic field $B_{||}$.}
\end{figure}
\cite{mani,zudov,studenikin} in two-dimensional electron system
(2DES). These striking results have been the subject of intense
current interest in the condensed matter community. Thus, new
experimental evidences are being published in a continuous basis
challenging the available theoretical models
\cite{ina2,girvin,dietel,lei,ryzhii,rivera,torres}. Among all of
those experimental outcomes we can cite for instance, temperature
dependence of MIRO\cite{mani,zudov,ina2}, absolute negative
conductivity \cite{willett,ina3,bykov}, bichromatic MW
excitation\cite{zudov2,ina4}, polarization immunity of
magnetoresistance ($\rho_{xx}$)\cite{smet,ina5,ng} and  the effect
of MW frequency on MIRO and ZRS\cite{studenikin2,ina6}. An important
set of results on MIRO that has not received yet special attention
from theorists is the effect of an in-plane magnetic field
($B_{||}$) (parallel to the 2DES, x-y plane) on the transport
properties of these devices. In a two-axis magnet experiment, we
have two magnetic fields: $B_{||}$ and the magnetic field
perpendicular ($B_{\bot}$) to the 2DES\cite{yang}. In a tilted
magnetic field ($B_{tilted}$) experiment, we only have one field and
in this case $B_{||}$ is the parallel component of
$B_{tilted}$\cite{mani2}. In the first case\cite{yang} (Yang's
experiment) results show an strong suppression of MIRO and ZRS. The
second case\cite{mani2} (Mani's experiment) demonstrates a
displacement of MIRO at larger $B_{tilted}$ and an unbalanced
quenching. Therefore these unexpected results deserve to be
considered by the different theoretical approaches serving as a
crucial test for them.

In this paper we present a theoretical model to address those
results and try to reconcile them using a common physical mechanism.
When a 2DES is illuminated with MW radiation,
electronic orbits are forced to move back and forth, oscillating
harmonically at the frequency of MW radiation and with an amplitude
proportional to the MW electric field\cite{ina2,ina3}. MIRO are
proportional to the magnitude of this amplitude and any variation of
it, is finally reflected on MIRO. In their MW-driven orbits motion,
electrons interact with the lattice ions being damped and producing
acoustic phonons. According to our model, the presence of $B_{||}$
imposes an extra harmonically oscillating motion in the
$z$-direction enlarging the electrons trajectory in their orbits.
This would increase the interactions with the lattice making the
damping process more intense  and reducing the amplitude of the
orbits oscillations. However depending on the origin of $B_{||}$,
the damping intensity would be different. Thus, in Yang's
experiment\cite{yang} the magnetic fields $B_{||}$ and $B_{\bot}$
are not related and $B_{||}$ is kept constant while $B_{\bot}$
increases from zero. Thus, the effect of $B_{||}$ is the same in the
whole $\rho_{xx}$ response and MIRO are uniformly  quenched. However
in Mani's experiment $B_{||}$ would not be constant because depends
on $B_{tilted}$. Then the quenching of MIRO would be unbalanced and
only visible at larger values of $B_{tilted}$.

\begin{figure}
\centering\epsfxsize=3.0in \epsfysize=2.3in
\epsffile{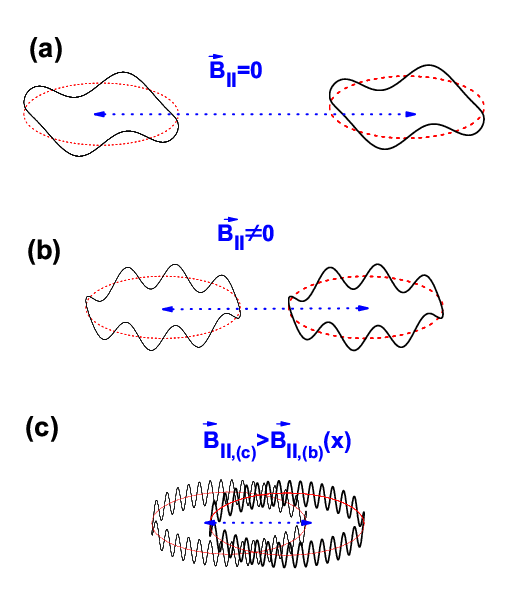} \caption{Schematic diagram showing the
dependence of MW-driven electronic orbit oscillating motion with
$B_{||}$. From (a) to (c), $B_{||}$ intensity increases which  makes
the MW-driven amplitude smaller and smaller. Eventually the
oscillating motion collapses, amplitude goes to zero and MIRO are
quenched. }
\end{figure}

%\section{The model}
The Hamiltonian for electrons confined in a 2D system (x-y plane) by
a potential $V(z)$ and subjected to a magnetic field
$B=(B_{x},0,B_{z})$, ($B_{||}=B_{x}$ and $B_{\bot}=B_{z}$) is given
by :
\begin{eqnarray}
&&H_{0}=\frac{P_{x}^{2}+P_{y}^{2}}{2m^{*}}+\frac{w_{z}}{2}L_{z}+\frac{1}{2}m^{*}\left[\frac{w_{z}}{2}\right]^{2}
(x^{2}+y^{2})\nonumber\\
&&+\frac{P_{z}^{2}}{2m^{*}}+\frac{1}{2}m^{*}w_{x}^{2}
z^{2}+V(z)+\frac{1}{2}w_{x}z(exB_{z}-2P_{y})\nonumber \\
&&=H_{xy}+H_{z}+ \frac{1}{2}w_{x}z(exB_{z}-2P_{y})
\end{eqnarray}
%where
%\begin{eqnarray}
%&&H_{xy}=\frac{P_{x}^{2}+P_{y}^{2}}{2m^{*}}+\frac{w_{z}}{2}L_{z}+\frac{1}{2}m^{*}\left[\frac{w_{z}}{2}\right]^{2}
%(x^{2}+y^{2}) \\
%&&H_{z}=\frac{P_{z}^{2}}{2m^{*}}+\frac{1}{2}m^{*}w_{x}^{2}
%z^{2}+V(z)
%\end{eqnarray}
We have used the symmetric gauge for $B_{z}$:
$\overrightarrow{A_{B_{z}}}=-\frac{1}{2}\overrightarrow{r}\times
\overrightarrow{B}=(-\frac{y}{2}B_{z},\frac{x}{2}B_{z},0)$, and the
Landau gauge for $B_{x}$:
$\overrightarrow{A_{B_{x}}}=(0,-zB_{x},0)$. $w_{z}$ is the cyclotron
frequency of $B_{z}$: $w_{z}=\frac{eB_{z}}{m^{*}}$ and  $w_{x}$ of
$B_{x}$: $w_{x}=\frac{eB_{x}}{m^{*}}$. $L_{z}$ is the z-component of
the electron total angular momentum. According to the  experimental
parameters used\cite{yang,mani2} the hamiltonian term  $
\frac{1}{2}w_{x}z(exB_{z}-2P_{y})<< H_{xy}+H_{z}$. Then, we can
discard this term and write:
%\begin{equation}
$H_{0}\simeq   H_{xy}+H_{z}$
%\end{equation}
%$ H_{xy}$ and $H_{z}$  are the Hamiltonians of two quantum harmonic
%oscillators, the first one is two-dimensional in the $x-y$ plane and
%the last one one-dimensional in the $z$ direction.

When one considers a parabolic potential for $V(z)$:
%\begin{equation}
$V(z)=\frac{1}{2}m^{*}w_{0}^{2} z^{2}$,
%\end{equation}
the Hamiltonian  $H_{z}$  can finally be written as:
\begin{equation}
H_{z}=\frac{P_{z}^{2}}{2m^{*}}+\frac{1}{2}m^{*}(w_{x}^{2}+w_{0}^{2})
z^{2}=\frac{P_{z}^{2}}{2m^{*}}+\frac{1}{2}m^{*}\Omega^{2} z^{2}
\end{equation}
and the Schrodinger equation of $H_{0}$ can directly be solved. We
obtain the wave functions of two harmonic oscillators, one is
two-dimensional in the $x-y$ plane and the other one is
one-dimensional in the $z$-direction.  In a semiclassical approach
the electron is subjected simultaneously to two independent harmonic
motions with a trajectory depicted in Figure 1: the electron
performs a circular movement in the $x-y$ plane and at the same time
a 1D harmonic oscillating motion in the $z$ direction. In Fig. 1(a)
we present the semiclassical trajectory of an electron in a 2D
parabolic potential: the electron trajectory is circular. 
In Fig.
%1(b), we add a parabolic potential in the $z$ direction, then the
%electron trajectory is circular in the plane and at the same time is
%oscillating in $z$. 
In Fig. 1(b) and 1(c),we add a parabolic potential in the $z$ direction, then the
electron trajectory is circular in the plane and at the same time is
oscillating in $z$.  We introduce also $B_{||}$ and
the oscillations in $z$ direction increase with the intensity of
$B_{||}$.  We obtain similar results when $B_{||}=B_{y}$.
\begin{figure} \centering\epsfxsize=3.0in \epsfysize=3.0in
\epsffile{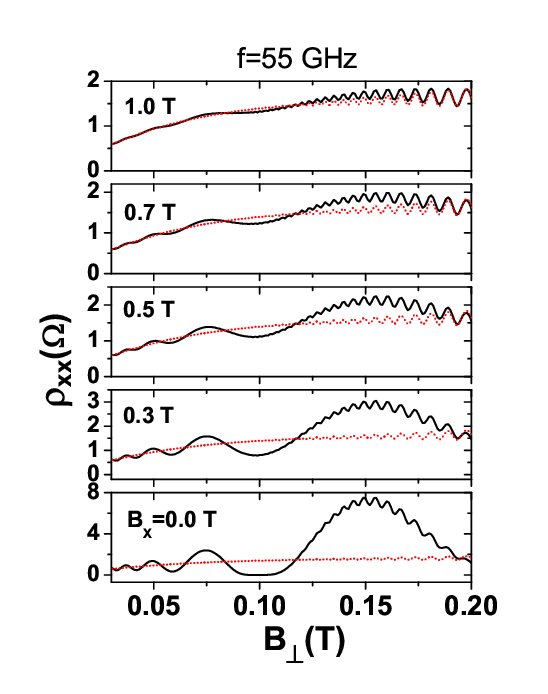}
\centering\epsfxsize=3.0in\epsfysize=2.5in
\epsffile{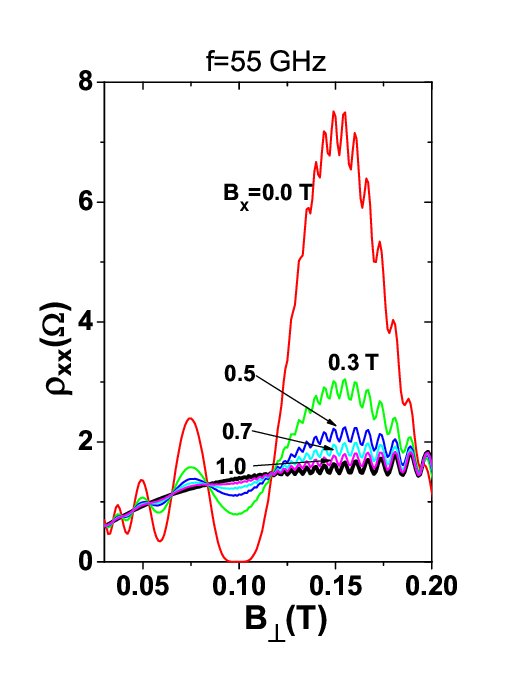} \caption{Calculated results of
$\rho_{xx}$ vs $B_{\bot}$ for different values of $B_{x}$ from $0.0$
T to $1.0$ T. $B_{\bot}$ and $B_{x}$ are independent. In the top
panel we present the curves of each value of $B_{x}$ in individual
panels. Single lines correspond with MW on, and dotted lines with MW
off. We observe a progressive quenching of $\rho_{xx}$ oscillations
as $B_{x}$ increases. In the bottom panel we present all curves
together for comparison. MW frequency is $55$ GHz. T=1K.}
\end{figure}

The problem of two-dimensional electrons subjected to magnetic
fields at arbitrary angles has been already solved analytically with
a parabolic confinement in the perpendicular
direction\cite{maan,merlin}. According to these results the
Hamiltonian can be transformed through a rotation of the coordinate
system with a certain angle. The obtained Hamiltonian corresponds to
two Hamiltonians of quantum harmonic oscillators and can be exactly
solved\cite{maan,merlin}.

If now we switch on the MW radiation (plane-polarized in the
$x$-direction) and connect a constant electric field ($E_{dc}$) in
the same direction (transport direction) the Hamiltonian then reads:
\begin{eqnarray}
&&H_{T}=\frac{P_{x}^{2}+P_{y}^{2}}{2m^{*}}+\frac{w_{z}}{2}L_{z}+\frac{1}{2}m^{*}\left[\frac{w_{z}}{2}^{2}\right]
\left[(x-X)^{2}+y^{2}\right]\nonumber\\
&&-\frac{1}{2}eE_{dc}X-eE_{0}(x-X) \cos wt -eE_{0}X\cos wt \nonumber\\
&&+\frac{P_{z}^{2}}{2m^{*}}+\frac{1}{2}m^{*}\Omega^{2} z^{2}
\end{eqnarray}
$X$ is the center of the orbit for the electron spiral motion:
%\begin{equation}
$X=\frac{eE_{dc}}{m^{*}(w_{c}/2)^{2}}$.
%\end{equation}
 The corresponding time-dependent Schrodinger equation can be exactly
solved\cite{ina2,ina3,ina4} and the wave functions are:
\begin{eqnarray}
\Psi_{T}(x,y,t)\propto\phi_{N}\left[(x-X-a(t)),(y-b(t)),t\right]
\phi(z)
%&&\times  \exp \frac{i}{\hbar} \left[m^{*}\left(\frac{d
%a(t)}{dt}x+\frac{d b(t)}{dt}y\right)+
%\frac{m^{*}w_{c}(b(t)x-a(t)y)}{2}-\int_{0}^{t} {\it L} dt'\right]\nonumber  \\
%&&\times\sum_{p=-\infty}^{\infty} J_{p}(A_{N}) e^{ipwt} \phi(z)\\
%&& = \Psi_{N} \phi(z)
\end{eqnarray}
where $\phi_{N}$ are
%the analytical solutions for the
%Schr\"{o}dinger equation with a 2D parabolic confinement, known as
Fock-Darwin states\cite{fock} and $\phi(z)$ is the one-dimensional
harmonic oscillator wave function.
%$J_{p}$ are Bessel functions with
%arguments, $A_{N}$\cite{ina2,ina3,ina4} and ${L}$ is the classical
%lagrangian.
$a(t)$ (for the x-coordinate) and $b(t)$ (for the y-coordinate) are
the solutions for a $classical$ driven 2D harmonic oscillator. The
expression, for instance, for $a(t)$\cite{ina5} is:\\
%\begin{equation}
$a(t)=\frac{e
E_{o}}{m^{*}\sqrt{(w_{z}^{2}-w^{2})^{2}+\gamma^{4}}}\cos wt=A\cos
wt$, where
%\end{equation}
\begin{figure}
\centering\epsfxsize=3.0in \epsfysize=3.0in
\epsffile{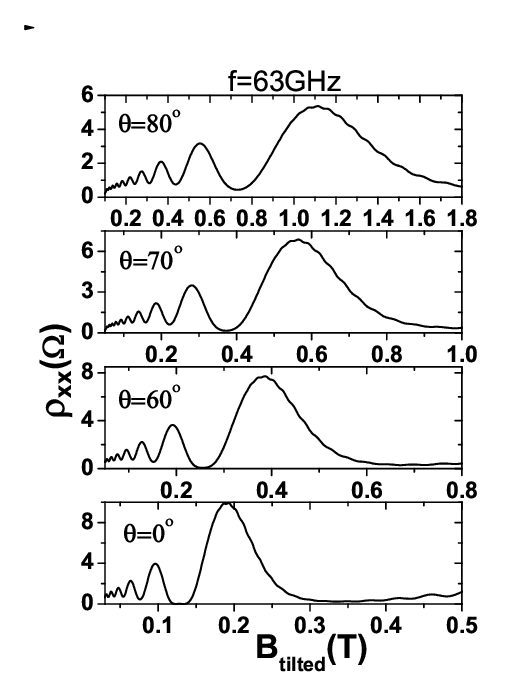} \centering\epsfxsize=3.0in
\epsfysize=2.5in \epsffile{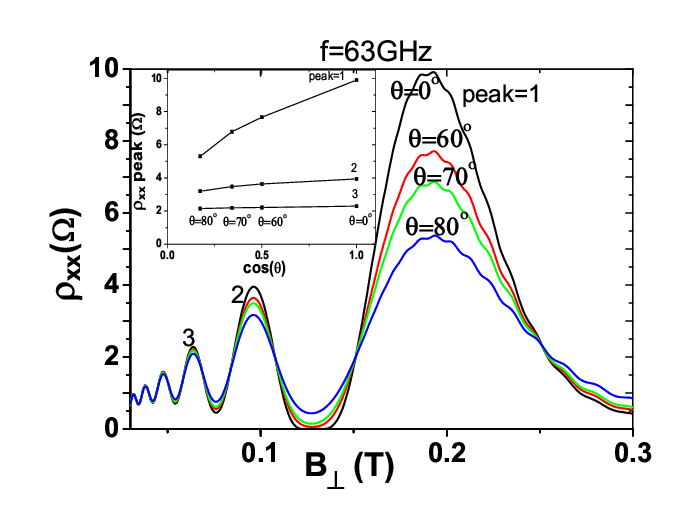} \caption{Calculated
results of $\rho_{xx}$ vs $B_{tilted}$ for different values of
$\theta$. In the top panel we present the curves of each value of
$\theta$ in individual panels. In the bottom panel we present all
curves together for comparison vs $B_{\bot}$. MW frequency is $63$
GHz. In the inset we present the peak resistance vs $\cos \theta$.
T=1K.}
\end{figure}
$E_{0}$ is the amplitude of the MW electric field, $w$ the frequency
and $e$ the electron charge. $\gamma$ is a damping factor which
dramatically affects the movement of the MW-driven electronic
orbits. Along with this movement interactions occur between
electrons and lattice ions, yielding acoustic phonons and producing
a damping effect in the electronic motion. In Ref.\cite{ina2,ina6},
we developed a microscopical model to calculate $\gamma$. We
obtained that $\gamma$ is a material and sample-dependent parameter
which depends also on lattice temperature, electronic orbit length
and MW frequency\cite{ina2,ina6}.
%Considering the ranges of lattice temperatures and
%MW frequencies used in experiments we have estimated an average
%value of $\gamma \sim 10^{12}s^{-1}$ for GaAs.
As we have indicated above, in a semiclassical explanation the
presence of $B_{||}$ alters the electron trajectory in its orbit
increasing the frequency and the number of oscillations in the
$z$-direction. Now the frequency of the $z$-oscillating motion  is
$\Omega>w_{0}$. This makes longer the electron trajectory increasing
the total orbit length and eventually the damping. This increase in
the orbit length is proportionally equivalent to the increase in the
number of oscillations in the $z$-direction. Thus, we introduce the
ratio of frequencies after and before connecting $B_{||}$ as a
correction factor for the damping factor $\gamma$. The final damping
parameter $\gamma_{f}$ is:
\begin{equation}
\gamma_{f}=\gamma \times \frac{\Omega}{w_{0}}=\gamma \times
\sqrt{1+\left(\frac{w_{x}}{w_{0}}\right)^{2}}= \gamma \times
\sqrt{1+\left(\frac{eB_{x}}{m^{*}w_{0}}\right)^{2}}
\end{equation}

%Now we introduce the scattering suffered by the electrons due to
%charged impurities randomly distributed in the
%sample\cite{ridley,ina2,ina3}. Firstly we calculate the
%electron-charged impurity scattering rate $1/\tau$ (being $\tau$ the
%scattering time), and secondly we find the average effective
%distance advanced by the electron in every scattering jump: $\Delta
%X^{MW}=\Delta X^{0}+ A\cos w\tau$, where $\Delta X^{0}$ is the
%effective distance advanced when there is no MW field present.
%Finally the longitudinal conductivity $\sigma_{xx}$ can be
%calculated: $\sigma_{xx}\propto \int dE \frac{\Delta
%X^{MW}}{\tau}(f_{i}-f_{f})$.
%being $f_{i}$ and $f_{f}$ the
%corresponding distribution functions for the initial and final
%states respectively and $E$ energy.
%To obtain $\rho_{xx}$ we use the relation
%$\rho_{xx}=\frac{\sigma_{xx}}{\sigma_{xx}^{2}+\sigma_{xy}^{2}}
%\simeq\frac{\sigma_{xx}}{\sigma_{xy}^{2}}$, where
%$\sigma_{xy}\simeq\frac{n_{i}e}{B}$ and $\sigma_{xx}\ll\sigma_{xy}$.
Now, following a previous model developed by us\cite{ina2,ina3}, we
are able to calculate $\rho_{xx}$, resulting that is proportional to
the MW-induced oscillation amplitude of the electronic orbits
center:
\begin{equation}
\rho_{xx}\propto A \cos
w\tau=\frac{eE_{o}}{m^{*}\sqrt{(w_{z}^{2}-w^{2})^{2}+\gamma_{f}^{4}}}\cos
w\tau
\end{equation}
%This expression shows that the MW-driven electron orbit amplitude
%$A$ will get smaller as $B_{=}$ increases (see Fig. 2) and
%eventually $\rho_{xx}$ will be quenched.
where $\tau$ is the charged impurity scattering
time\cite{ina2,ina3}.
%\section{Results}
In Figure 3 we present calculated results of $\rho_{xx}$ vs
$B_{\bot}$ for different values of $B_{x}$ being $B_{x}$ independent
of $B_{\bot}$ as in Yang's experiment. Thus, calculations have been
made for a MW frequency of  $55$ GHz and a confinement in $z$  of
$50$ nm (similar as Yang's experiment). $B_{x}$ values are:
$B_{x}(T)=0.0$, $0.3$, $0.5$, $0.7$ and $1.0$. In the top panel we
can see the results for each value of $B_{x}$ in individual panels.
We observe clearly the progressive quenching of MIRO as $B_{x}$
increases. In the bottom panel we present all curves together in the
same panel for comparison. According to equations (5) and (6), when
we increase $B_{x}$, we also increase  the damping $\gamma_{f}$ and
as a result the amplitude $A$ and MIRO are progressively quenched.
We obtain a total quenching for $B_{x}\simeq 1.0$ T. We observe a
uniform damping in the whole range of $B_{\bot}$ for each value of
$B_{x}$. These results are in good agreement with
experiment\cite{yang}. In Figure 4 we present calculated results of
$\rho_{xx}$ vs $B_{tilted}$ in the top panel and vs
$B_{\bot}=B_{tilted} \cos \theta$ in the bottom panel. According to
parameters and set-up of Mani's experiment, we have used a
confinement in $z$ of $30$ nm and MW frequency of $63$ GHz. We
present different curves corresponding to different tilt angles
($\theta$). $\theta$ values are: $\theta=0^{0}, 60^{0}, 70^{0},
80^{0}$. In this case $B_{||}=B_{tilted} \sin \theta$. In the top
panel we present calculated curves for each $\theta$ in individual
panels. We observe MIRO displacement at larger $B_{tilted}$ for
increasing values of $\theta$. This is explained considering that
$B_{\bot}=B_{tilted}\cos \theta$ and that peak positions are only
governed by $B_{\bot}$. In the bottom panel we can see all curves
together for comparison.  We observe, as in experiment, how the
quenching effect is not uniform and  more intense with increasing
values of $B_{tilted}$ and $\theta$. Our model explains this
peculiar behavior with the expression of $B_{||}$ and equations (5)
and (6). Increasing values of $B_{tilted}$ and $\theta$ give rise to
an increasing damping according to equation (5) and decreasing $A$
and MIRO according to equation (6). Thus, we observe a soft
quenching for small values of $B_{\bot}$ and $\theta$, and it gets
progressively more important for larger values of  them. This
feature is shown in the inset of Fig. 4 bottom panel, where we
present the peak resistance vs $\cos\theta$  for peaks$=1,2,3$, as
in experiment\cite{mani2}. The decrease in the peak resistance is
larger for peak $1$ than for peaks $2,3$ as $\theta$ increases.
These results are in good agreement with Mani's
experiment\cite{mani2}.

% In the case of $B_{=}$
%constant and independent of  perpendicular $B$ (Yang's experiment),
%the $A$ damping and $\rho_{xx}$ quenching will be the same for all
%values of $B$ affecting similarly all oscillations. For a certain
%value of $B_{=}$, $\rho_{xx}$ oscillations will be totally quenched
%. However in the case of a tilted $B$ (Mani's experiment), $B_{=}=B
%\sin \theta$ (being $\theta$  the angle of inclination) and then
%damping and quenching are small for small values of $B$ and increase
%as $B$ increases. Then we expect that the damping effect of  $B_{=}$
%will be less effective in the case of

%\section{Conclusions}
In summary, we have presented a theoretical model on the effect of an
in-plane magnetic field on MIRO and ZRS in 2DES. Experimental
results show different behaviors depending on the set-up of
the magnetic field. In an independent $B_{||}$ experiments report a clear
quenching of MIRO and ZRS. In a tilted $B$, experiments show
oscillations displacement and an unbalanced quenching. We have
presented a theoretical model which explains these results based
in a common physical mechanism. The MW-driven oscillating electronic
orbit motion is increasingly damped by the presence of $B_{||}$. The understanding
of this behavior will allow to control the transport properties in a MW
irradiated Hall bar. In particular MIRO and ZRS features by tuning external
magnetic fields in different configurations.

This work has been supported by the MCYT (Spain) under grant
MAT2005-0644 and by the Ram\'on y Cajal program.

\end{document}